 \documentclass[twocolumn,aps,prd,floats,floatfix,showpacs,superscriptaddress,nofootinbib]{revtex4}
\usepackage{amsmath,amsfonts}\usepackage{graphicx}
\begin{document}
\title{The general spherically symmetric constant mean curvature foliations of the Schwarzschild solution.}
\author{Edward Malec}\affiliation{Physics Department, Jagiellonian University, Krakow, Poland}
\affiliation{Physics Department, University College, Cork, Ireland}\author{Niall \'{O} Murchadha}
\affiliation{Physics Department, Jagiellonian University, Krakow, Poland}\affiliation{Physics Department,
 University College, Cork, Ireland}
\begin{abstract}
We consider a family of spherical three dimensional spacelike
 slices embedded in the Schwarzschild solution. The mean curvature is constant
 on each slice but can change from slice to slice.    We give a simple expression for an everywhere positive lapse and thus we show how to construct foliations. There is a barrier
preventing the mean curvature from becoming large, and we show how to avoid this so as to construct a foliation
 where the mean curvature runs all the way from zero to infinity. No foliation exists 
where the mean curvature goes from minus to plus infinity. There are
  slicings, however,  where each slice passes through the bifurcation sphere 
  $R = 2M$ and  the lapse only vanishes
 at this one point, and is positive everywhere else, while the mean curvature does 
run from minus to plus infinity. Symmetric foliations of the extended Schwarzschild
 spacetime degenerate at a critical point, where we show that 
the lapse function exponentially approaches  zero.
\end{abstract}
\maketitle

\section{Introduction}

Constant mean curvature (CMC) foliations of the Schwarzschild geometry have been
constructed by Brill, Cavalho and Isenberg \cite{Brill}. These foliations degenerate 
when the lapse ``collapses"; foliations in the vicinity of 
this critical point have been  been investigated in
\cite{MOM}. In both of these papers, it is assumed that the trace of the extrinsic curvature is not only a constant on each slice, but retains this constant value from slice to slice. In this paper we analyse   more general families of CMC foliations, with the 
  trace $K$ of the extrinsic curvature that  
  can change with time.
The   ``collapse" of the lapse  can again be described by analytic approximations.

The standard way of viewing General Relativiy as a dynamical system is by considering 
the 4-manifold as foliated by a sequence of spacelike 3-surfaces  \cite{ADM, MTW}. Each 3-surface inherits 
a 3-metric, $g_{ij}$, and an extrinsic curvature $K_{ij} =  (1/2) {\mathcal L}_{\bf n} g_{ij}$ \cite{W} 
where ${\mathcal L}_{\bf n}$ is the Lie derivative along the normal, and thus $K_{ij}$ is a geometric 
object, the analogue of the time derivative of the metric. Each slicing is equivalent to a choice 
of time function. A standard choice is to demand that the trace of the extrinsic curvature, $g^{ij}K_{ij}$, 
usually written as $K$ and known as the mean curvature of the surface, be constant along each slice. 
Hence these are the constant mean curvature, or CMC, slices.CMC slices are attractive for a number 
of reasons. The Einstein equations give an evolution equation for $K$, it is (in vacuum)
\begin{equation} \nabla^2 N - K^{ij}K_{ij} N = N{\mathcal L}_{\bf n} K =
 {\partial K \over \partial t} -N^i\partial_i K,\end{equation}
where $N$ and $N^i$ are the lapse and shift, respectively.

If $K$ is a spatial constant, the shift term drops out and the equation reduces to
\begin{equation} \label{eq:dK/dt}\nabla^2 N - K^{ij}K_{ij} N = {\partial K \over \partial t}.\end{equation}
 and this can be regarded as an equation for the lapse function, $N$, of a CMC slicing.
 This is a nice linear elliptic equation that satisfies the maximum principle. 
 
 In the standard, conformal, methods of constructing initial data for the gravitational field, 
choosing the trace of the extrinsic curvature to be a constant simplifies the equations. 
One gets a single, nonlinear scalar equation for the conformal factor instead of 
a coupled system of nonlinear equations \cite{GC}.
 
 By taking the trace of the equation defining $K_{ij}$, one can show ${\mathcal L}_{\bf n} \sqrt{g} = \sqrt{g} K$. 
This tells us that $K$ is just the fractional time rate of change of the volume along the normal, and, 
in a cosmology, the CMC slices are the `Hubble time' slices, with the instantaneous value of $K$ 
equalling the Hubble `constant'.In Minkowski space, on the other hand, the `mass hyperboloids', 
$t^2 - r^2 = m^2$, are CMC slices, with $K = 3/m$.  Slices in general asymptotically flat spacetimes
 mimic the behaviour of the mass hyperboloids, in that they are everywhere spacelike but become null
 at null infinity. This feature makes them   of value for the 
analysis of gravitational or any other form of radiation, see e.g. \cite{AZ} \cite{O}.  
 
 One major advantage of CMC slicings for numerical relativity is that, 
if we consider gravitational waves of a fixed wavelength, 
the number of wave cycles that a CMC slice intercepts is finite. To resolve a wave in a numerical computation,
 we need the separation between data points to be less than the wavelength. This means that the domain of a code can extend 
all the way to null infinity, and track the waves all the way out with a finite number of grid points. 
This phenomenon is independent of whether one compactifies or not.
 
 \section{Explicit CMC foliations}We possess a great deal of understanding about the spherical CMC slices 
of the Schwarzschild solution \cite{MOM}. We can write the 3-metric and extrinsic curvature analytically 
in terms of the Schwarzschild radius $R$. They depend only on two parameters, $K$, the trace of 
the extrinsic curvature, and $C$, that quantifies the transverse-traceless part of the extrinsic curvature. 
The expressions are
\begin{equation} \label{eq:g_{ij}}dS^2 = {dR^2 \over 1 - {2M \over R} + \big( {KR \over 3} - {C \over R^2}\big)^2} + R^2 
d\Omega^2.\end{equation}
The extrinsic curvature is diagonal, with the non-zero components being
\begin{equation} \label{eq:K^i_j}K^R_R = {K \over 3} + {2C \over R^3}; K^\theta_\theta = K^\phi_\phi = 
{K \over 3} - {C \over R^3}.\end{equation}
The sixth-order polynomial
\begin{equation} \label{eq:k^2}k^2 = 1 - {2M \over R} + \left( {KR \over 3} - {C \over R^2}\right)^2\end{equation}
plays a key role. It has several meanings, $k = dR/dL$, where $L$ is the proper distance along 
the slice, therefore $2k/R$ is the 2-mean- curvature of the round 2-spheres as 
embedded in the 3-slice.  Also $k = N_K$, where $N_K$ is the Killing lapse, the dot product of 
the `timelike' Killing vector with the unit normal to the slice. For small values of $K$ and $C$, $k^2$ 
is positive for both small and large $R$, with only two roots.  We must have positive $k^2$ for 
the metric to make physical sense. Let us label the larger of these two roots $R_t$. One can show easily that 
$R_t < 2M$. The zone where $k^2$ is positive from $R = R_t$ to $R = \infty$ corresponds to a slice 
that starts out at one future null infinity (if $K > 0$), reaches a minimum surface at $R = R_t$ 
(hence $`t'$ for `throat') and continues on to the other future null infinity.  If $C < 8M^3K/3$ it crosses 
below the bifurcation point, if $C > 8M^3K/3$ it crosses through the upper quadrant. As $C$ becomes larger, 
while holding $K$ fixed, the polynomial $k^2$ rises up, the two roots approach each other until they finally 
touch at a value of $R$ which we call $R_\star$. The value of $K$ and $C$ at that point equal
\begin{equation} \label{eq:Kstar}K_\star = {2R_\star - 3M \over \sqrt{2MR^3_\star - R^4_\star}}; C_\star = 
{3MR^3_\star - R^4_\star \over 3\sqrt{2MR^3_\star - R^4_\star}}.\end{equation}
This is a critical point of the slicing, As $(C, K)$ change so as to approach this point one gets 
the `collapse of the lapse' and `slice stretching' that one is used to in maximal slicing\cite{BOM, MOM}. 
If we increase $C$ even further, $k^2$ is positive over the entire range $R = (0, \infty)$. When this is true,  
the CMC slice starts at null infinity and plunges into the singularity.We have previously analysed the CMC slices 
where we hold $K$ fixed and just allowed $C$ to change \cite{MOM}. Obviously there exists a much richer class 
of CMC slicings, where both $C$ and $K$ change. One really only needs to give some relationship between $C$ and $K$. 
However it is easier to introduce some parameter time $t$ and write $C(t)$ and $K(t)$. One can always reparametrise, 
which will change the form of both $C(t)$ and $K(t)$. However, the ratio, $(dK/dt)/(dC/dt)$, is unchanged.
 
 The $G_{RR}=0$  Einstein equation can be written as 
\begin{equation}
\partial_t\left( RK^\theta_\theta \right) |_{R=const}=k^3\partial_R {N\over k}.
\label{Grr}
\end{equation}

This equation  is solved by  the   lapse function $N $, where

\begin{equation}
N \equiv   \beta k +  k\int_{R}^{\infty }dr{\dot C-{r^3\over 3} 
\dot K\over  r^2k^3} .
\label{lapse}
\end{equation}

Here $\beta $ is a (time-dependent) constant.
One can verify that the lapse equation (\ref{eq:dK/dt}), is solved by $N $ as given in Eq.(\ref{lapse}).

It is clear that Eq.(\ref{eq:dK/dt}) is a linear elliptic inhomogeneous equation, so therefore the solution
 can be written as a linear combination of a `particular' solution of the inhomogeneous equation combined 
with a solution of the homogeneous equation $(\nabla^2 - K^{ij}K_{ij})\psi = 0$. Since $k$ is the Killing 
lapse, we know that 
\begin{equation}
(\nabla^2 - K^{ij}K_{ij})k = 0. 
\label{k}
\end{equation}
We also know that we can change $C$ without changing $K$. This also gives us a lapse that satisfies
 the homogeneous equation, call it $\phi(\dot C)$. It can be chosen to depend linearly on $\dot C$ 
(one may have to subtract off some multiple of $k$). Finally, the inhomogeneous solution can be can 
be adjusted so as to linearly depend on $\dot K$ (again, one may have to subtract off multiples of $k$ 
and $\dot C$). This is why we get the three constants $\beta$, $\dot C$, and $\dot K$ appearing 
explicitly in Eq.(\ref{lapse}).

Let us now restrict ourselves to the case where $C$ and $K$ are small enough that $k$ has a zero at 
$R = R_t$. Let us also assume $K > 0$. This means that the slice comes in from one future null infinity, 
has a minimal surface at $R = R_t$ and goes out to the other future null infinity. We know that 
the Killing lapse $k$, the solution to Eq.(\ref{k}), is anti-symmetric, passing through zero at $R = R_t$. 
By adding an appropriate multiple of $k$ we can find  the solution that is proportional to $\dot K$ 
and the solution proportional to $\dot C$ that are symmetric around the throat.

Let us have a spherical CMC slicing of the Schwarzschild solution. This can be described as 
a curve in $(C, K)$ space. There is still some freedom.
The term in the lapse proportional to $k$ can be thought of as `pure gauge'. It slides the slice 
along the Killing vector without changing either $K$ or $C$. Since we are interested in 
the effects of changing $K$ and $C$ we would like to eliminate this freedom.  
The obvious way is to adjust the lapse by adding an appropriate amount of $k$ so that 
the lapse is symmetric around the throat. This gives us a Neumann boundary condition 
$dN/dL = dR/dL dN/dR = k dN/dR = 0$ at $R = R_t$ where $L$ is the proper distance along the slice. 
Since $k = 0$ at the throat, this looks like a trivial equation.  It is not, because 
$dN/dR$ generically blows up at the throat like $1/k$. This has the effect of reducing the number 
of constants in the solution from three to two. The second freedom is the choice of time 
parametrisation of the slicing. It appears that the `natural' choice is to set $\beta = 1$ in 
Eq.(\ref{lapse}). This allows the time translation vector of the slicing to coincide with 
the timelike Killing vector at null infinity. It turns out that we can satisfy both conditions, 
i.e., $\beta = 1$ and the symmetry at the throat, simultaneously. 

It is interesting to look at expression Eq.(\ref{lapse}) in the limit of large $R$. 
In that limit $k \approx KR/3$. Therefore the integrand is dominated by $-9\dot K/K^3 r^2$. 
This integrates to $9\dot K/K^3 r$, so the integral equals $-9\dot K/K^3 R + O(1/R^2)$. 
The $k$ outside the integral becomes $KR/3$ so we finally get
\begin{equation}
N \approx \beta k - {3\dot K \over K^2} + O(1/R).
\end{equation}
Therefore $N/k \rightarrow \beta$ for large $R$. Since $k$ is the Killing lapse, if $\beta = 1$ 
we can have the time translation of the slicing equal the Killing vector. 
With this choice the time parameter is the retarded time.

While expression Eq.(\ref{lapse}) is easy to understand near infinity, it is difficult to see from 
it what happens near the throat. 
To find the cleanest expression, we have to manipulate Eq.(\ref{lapse}). It is easy to see that 
\begin {equation}
{d \over dR}\left({1 \over k}\right) =  - {1 \over k^3}\left({M \over R^2} + {K^2 R\over 9} 
+ {KC \over 3R^2} - {2C^2 \over R^5}\right).
\end{equation}
Therefore we have
\begin{equation}
{1 \over r^2k^3} = - \left( {1 \over M + {KC \over 3} + {K^2r^3 \over 9} - {2C^2 \over r^3}}\right) 
{d \over dr}\left({1 \over k}\right),
\end{equation}
and the expression for the lapse can be rewritten as
\begin{equation}
N = \beta k - k\int_R^{\infty}\left( {\dot C - {\dot K r^3\over 3} \over M + {KC \over 3} + 
{K^2r^3 \over 9} - {2C^2 \over r^3}}\right) {d \over dr}\left({1 \over k}\right)dr.
\end{equation}
Now integrate by parts to get
\begin{eqnarray}
N &=& \beta k + k\int_R^{\infty}{1 \over k}{d \over dr}\left( {\dot C - {\dot K r^3\over 3} 
\over M + {KC \over 3} + {Kr^3 \over 9} - {2C^2 \over r^3}}\right)dr \nonumber\\ &-& \left. 
k\left({1 \over k} {\dot C - {\dot K r^3\over 3} \over M + {KC \over 3} + {K^2r^3 \over 9} 
- {2C^2 \over r^3}}\right) \right|_R^{\infty}.
\end{eqnarray}
If we change the range of integration from the throat to $R$, instead of from $R$ to $\infty$ we get
\begin{eqnarray}
N &=& \beta k + k\int_{R_t}^{\infty}{1 \over k}{d \over dr}\left( {\dot C - {\dot K r^3\over 3}
 \over M + {KC \over 3} + {Kr^3 \over 9} - {2C^2 \over r^3}}\right)dr \nonumber\\ 
&-& k\int_{R_t}^R{1 \over k}{d \over dr}\left( {\dot C - {\dot K^2 r^3\over 3} \over M + {KC \over 3} 
+ {Kr^3 \over 9} - {2C^2 \over r^3}}\right)dr \nonumber\\ 
&+&   {\dot C - {\dot K R^3\over 3} \over M + {KC \over 3} + {K^2R^3 \over 9} - {2C^2 \over R^3}} .
\end{eqnarray}
It turns out that if we choose 
\begin{equation}\label{beta}
\beta  = -\int_{R_t}^{\infty}{1 \over k}{d \over dr}\left( {\dot C - {\dot K r^3\over 3} \over M + 
{KC \over 3} + {Kr^3 \over 9} - {2C^2 \over r^3}}\right)dr,
\end{equation}
i.e., eliminate the term that is of the form constant $\times k$ we get the symmetric lapse function
  that satisfies $dN/dL = 0$ at the throat. This is
\begin{eqnarray} \label{eq:N}
N = 
&-& k\int_{R_t}^R{1 \over k}{d \over dr}\left( {\dot C - {\dot K r^3\over 3} \over M + {KC \over 3} + 
{Kr^3 \over 9} - {2C^2 \over r^3}}\right)dr \nonumber\\ 
&+&   {\dot C - {\dot K R^3\over 3} \over M + {KC \over 3} + {K^2R^3 \over 9} - {2C^2 \over R^3}} .
\end{eqnarray}We obviously assume that we are in the subcritical regime so that $R_t$ exists.  
It is straightforward, if tedious, to show that this expression satisfies the CMC equation, Eq.(\ref{eq:dK/dt}). 

As stated earlier, we can simultaneously set $dN/dL = 0$ at the throat (just by choosing the lapse 
as given by Eq.(\ref{eq:N})) and simultaneously set $\beta = 1$. We start off with a curve in 
$(C, K)$ space on which we put some coordinate time label. This defines $\dot C$ and $\dot K$. 
Now compute 
\begin{equation}\label{gamma}
\gamma  = \int_{R_t}^{\infty}{1 \over k}{d \over dr}\left( {-\dot C + {\dot K r^3\over 3} 
\over M + {KC \over 3} + {Kr^3 \over 9} - {2C^2 \over r^3}}\right)dr,
\end{equation}
and scale the time by $\gamma$, i.e.,
\begin{equation}
d\hat t = \gamma dt.
\end{equation}
This rescales $\dot C$ and $\dot K$ by $\gamma$.  Now the value of $\beta$ for which the two 
lapse functions, Eq.(\ref{lapse}) and Eq.(\ref{eq:N}), are equal is $\beta =1$!  

The quantity $M + K^2r^3/9 + KC/3 -  2 C^2/r^3$ that appears twice in the expression for $N$ is
 nothing but ($r^2/2$ times) the first derivative of $k^2$ and this is positive on the entire 
interval $[R_t, \infty)$. It only goes to zero as we approach the critical point. It is the `going to zero' 
of this function that gives the classical `collapse of the lapse'. 

This lapse, while maintaining the CMCness of the slices, will not keep the metric in the form Eq.
(\ref{eq:g_{ij}}), because the normal to the slice is not along the $R$ = constant direction.
 It is easy to work out what the appropriate shift is since the Killing vector is along
 this direction. We know that the Killing lapse is $\alpha_K = k$ and we also know that 
$\alpha_K^2 - \beta_K^2 = 1 - 2M/R$. Therefore $\beta^2_K = (KR / 3 - C /R^2)^2 \Rightarrow \beta^R_K = k(-KR / 3 
+ C /R^2)$. In turn we get $N^R = N(-KR / 3 + C /R^2)$. Since $g_{RR} = 1/k^2$ we 
get $N_R = N(-KR / 3 + C /R^2)$. This argument works for both choice of lapse,
 Eq.(\ref{lapse}) and Eq.(\ref{eq:N}). In the upper half plane, if $N > 0$, 
we want $N^R < 0$ for large $R$ because the normal to the CMC slice is 
leaning over more towards null infinity than the Killing vector is. 
This shift does not vanish at the throat for the symmetric $N$ as 
distinct from the Killing shift.  It will be positive (if $N$ is positive) 
there and we seek a symmetric solution. This means that we have a discontinuity  
in the shift across the throat. This is to be expected, because, as we move forward 
in time,  the coordinate range of $R$ expands.  The opposite will occur in 
the lower half plane.From the formula for the symmetric $N$, Eq.(\ref{eq:N}),  
we can easily read off the value of the lapse at the throat as
\begin{equation} \label{eq:throat}N(R = R_t) = {{dC \over dt} - {dK \over dt}{R_t^3 \over 3} \over M + 
{K^2R_t^3 \over 9} +{KC \over 3} - {2 C^2 \over R_t^3}}.\end{equation}
This is obviously positive if $dC/dt - dK/dt(R_t^3/3) > 0$. If we have a symmetric slicing 
and if $N = 0$ at any one point then it must be zero on an entire closed 2-sphere. 
If $dK/dt$ is positive, we can immediately see from applying the maximum principle 
to the CMC equation, Eq.(\ref{eq:dK/dt}), that if the lapse is positive in the center
 it is positive everywhere. One way to show this is to assume the opposite. 
 Let us assume that we have a region on the  3-surface where $N > 0$  in the interior, 
and $N = 0$ on the boundary. Therefore $dN/dR  \le 0$ on the boundary. Multiply 
the lapse equation, Eq.(\ref{eq:dK/dt}), by $N$ and integrate over the region in which 
$N \ge 0$. One gets an immediate contradiction!  Therefore we have a foliation instead 
of just a slicing. This kind of argument was first introduced by \cite{BOM}. 
This expression for the central lapse in Eq.(\ref{eq:N}) gives us an immediate consistency check. 
Let us consider the situation where the variations in $C$ and $K$ are such that the radius 
of the throat does not change. Since $R = R_t$ is the zero of $k^2$, by inspecting the formula 
for $k^2$, Eq.(\ref{eq:k^2}), the condition is that
\begin{equation}{\delta K R_t \over 3} - {\delta C \over R_t^2} = 0.\end{equation}
But, of course, if the radius does not change the lapse at the throat must vanish. 
This is exactly what we get from Eq.(\ref{eq:throat}).

\section{Two   spacetime metrics}
Let us start with a CMC slice whose 3-metric is given by \ref{eq:g_{ij}}, 
and let us drag this along the timelike Killing vector, i.e., we set $\dot K = 0$, $\dot C = 0$. 
We can now write down the 4-metric associated with this slicing (it is not a foliation 
because the lapse vanishes at the throat). It is
\[ \bar g_{\mu \nu}= \left(
\begin{array}{cccc}
 -\left(1 - {2M \over R}\right), &  {1\over k}\left({C \over R^2} - {KR \over 3} \right)
  & 0 & 0 \cr 
  {1\over k}\left({C \over R^2} - {KR \over 3}\right) , & {1 \over k^2} & 0 & 0\cr
  0 & 0 & R^2 & 0 \cr
   0 & 0 & 0 & R^2\sin^2\theta  \end{array}\right) \] 
 
 \[ \bar g^{\mu
 \nu}= \left( \begin{array}{cccc} -{1 \over k^2}, &   {1\over k}\left({C \over R^2} - {KR \over 3}\right)  & 0 & 0\cr
  {1\over k}\left({C \over R^2} - {KR \over 3} \right), & \left(1 - {2M \over R}\right) & 0 & 0
 \cr 0 & 0 & {1 \over R^2} & 0 \cr 0 & 0 & 0 & {1 \over R^2\sin^2\theta}\end{array}
  \right) \]
Another spacetime metric can be constructed using the same 3-metric but choosing a lapse function 
that allows $C$ and $K$ to change with time, i.e., either Eq.(\ref{lapse}) or Eq.(\ref{eq:N}). 
This has the form
\[ \bar g_{\mu \nu}= \left(
\begin{array}{cccc}
 -{N^2 \over k^2}\left(1 - {2M \over R}\right), & {N \over k^2}\left({C \over R^2} - {KR \over 3}\right)
  & 0 & 0 \cr 
 {N \over k^2}\left({C \over R^2} - {KR \over 3}\right), & {1 \over k^2} & 0 & 0\cr
  0 & 0 & R^2 & 0 \cr
   0 & 0 & 0 & R^2\sin^2\theta  \end{array}\right) \] 
 
 \[ \bar g^{\mu
 \nu}= \left( \begin{array}{cccc} -{1 \over N^2}, &   {1\over N}\left({C \over R^2} - {KR \over 3}\right)  
& 0 & 0\cr
 {1\over N}\left({C \over R^2} - {KR \over 3}\right) , & k^2 & 0 & 0
 \cr 0 & 0 & {1 \over R^2} & 0 \cr 0 & 0 & 0 & {1 \over R^2\sin^2\theta}\end{array}
  \right) \]

  \section{Matching CMC to maximal slicing}
  Using the explicit form of the metric(s) in the previous section, we can show that one can smoothly match 
a spherical CMC slicing (with either a constant or time-dependent $K$) to a maximal slicing in an extended Schwarzschild solution.  The matching is performed along a timelike surface with fixed Schwarzschild radius $R = R_j > 2M$. We can handle the case where the CMC slices are on the `inside' and the maximal slices are `outside', and the converse.
  
  We wish to match a patch of spacetime with metric
  \[ g_{\mu \nu}= \left(
\begin{array}{cccc}
 -{N^2 \over k^2}\left(1 - {2M \over R}\right), & {N \over k^2}\left({C \over R^2} - {KR \over 3}\right)
  & 0 & 0 \cr 
 {N \over k^2}\left({C \over R^2} - {KR \over 3}\right), & {1 \over k^2} & 0 & 0\cr
  0 & 0 & R^2 & 0 \cr
   0 & 0 & 0 & R^2\sin^2\theta  \end{array}\right) \] 
   with
   \begin{equation}
{N \over k} \equiv   \beta_K(t)  +  \int_{R}^{\infty }dr{\dot C-{r^3\over 3} 
\dot K\over  r^2k^3} 
\label{lapse1}
\end{equation}
along some cylinder of constant Schwarzschild radius $R = R_j$ to a patch of spacetime with metric
 \[ \tilde g_{\mu \nu}= \left(
\begin{array}{cccc}
 -{\tilde N^2 \over \tilde k^2}\left(1 - {2M \over R}\right), & {\tilde N \over \tilde k^2}{\tilde C \over R^2}
  & 0 & 0 \cr 
 {\tilde N \over \tilde k^2}{\tilde C \over R^2} , & {1 \over \tilde k^2} & 0 & 0\cr
  0 & 0 & R^2 & 0 \cr
   0 & 0 & 0 & R^2\sin^2\theta  \end{array}\right) \] 
    with
   \begin{equation}
{\tilde N \over \tilde k} \equiv   \beta_{\tilde C}(t)  +  \int_{R}^{\infty }dr{\dot {\tilde{C}}
\over  r^2\tilde k^3}, 
\label{lapse2}
\end{equation}
with $\tilde k^2 = 1 - {2M \over R} + {\tilde C^2 \over R^4}$. 

We need to check that the Israel-Darmois junction conditions \cite{Israel} are satisfied. Note that
 we allow $\beta$ to be two different time dependent functions, one in each patch. Really, 
we only need $\beta_K(t)$ to change with time, it makes sense to keep $\beta_{\tilde C} = 1$. 
This guarantees that the time in the `maximal' zone is the proper time at infinity. 
The Israel-Darmois junction conditions are that intrinsic metric and the extrinsic curvature 
of the 3-dimensional matching surface be continuous.  
We can actually arrange that the whole 4-metric be continuous.

The key condition is that we choose $\tilde C, K$ and $C$ to satisfy
 $\tilde C = C - {KR^3_j \over 3}$. This has to be satisfied on every time slice. Therefore we also 
want $\dot {\tilde C} = \dot C - {\dot KR^3_j \over 3}$.  This guarantees that $k = \tilde k$ 
along the matching surface. We also can use the free parameter $\beta_K(t)$ to maintain $\tilde N = N$. 
This guarantees that the 4-metrics to the left and right of the surface $R = R_j$ are the same.

Now we need to look at the extrinsic curvature. The condition $\tilde C = C - {KR^3_j \over 3}$ 
guarantees that $\tilde K_{\theta\theta} = K_{\theta\theta}$ and $\tilde K_{\phi\phi} = K_{\phi\phi}$.
 This means that we only need to check $K_{00}$. The normals to the surface $R = R_j$ 
are respectively $\tilde n_{\mu} = (0, {1 \over \tilde k}. 0, 0$ and $ n_{\mu} = (0, {1 \over  k}. 0, 0$. 
Hence $\tilde n_{\mu} = n_{\mu}$. We have  $K_{00} = n_{0;0} = -\Gamma^R_{00} n_R =
 - {1 \over 2}g^{R\mu}(2g_{0\mu,0} - g_{00,\mu})n_R$. Since the metrics match on the surface, 
we have $g_{0\mu,0} = \tilde g_{0\mu,0}$. Therefore the problem reduces to comparing 
$(\tilde N/\tilde k)_{,R}$ to $(N/k)_{,R}$. Looking at Eqns.(\ref{lapse1}) and (\ref{lapse2}). 
This reduces to comparing $\dot {\tilde C}$ to $ \dot C - {\dot KR^3_j \over 3}$. These are obviously equal. 
 Therefore we can match both the intrinsic metric and the extrinsic curvature along the 3-surface defined by $R = R_j$. 
 
 The metric along the matching surface is  $C^{\infty}$ while the matching of the extrinsic curvatures guarantees that the metric perpendicular to the surface is $C^1$. Further, consider the Hamiltonian and momentum constraints along the matching surface. The behaviour of the metric and extrinsic curvature means that there is no jump discontinuity in $^{(3)}R  - K^{ab}K_{ab} - K^2$ or $\nabla_a(K^{ab} - g^{ab}K)$ across the surface. This means that there are no stresses along the matching surface.
 
 Consider the Hamiltonian and momentum constraints on a spacelike CMC slice. These are $^{(3)}R - K^{ab}K_{ab} + K^2 = 0$ and $\nabla_a(K^{ab} - g^{ab}K) = 0$. If we write $K^{ab} = K^{ab}_t + 1/3 Kg^{ab}$ we can rewrite the constraints as $^{(3)}R - K^{ab}_tK^t_{ab} = -(2/3) K^2 $ and $\nabla_aK^{ab}_t = 0$. Therefore the initial data can be regarded as maximal data coupled to a constant negative-density `matter' field at rest with $16 \pi \rho = -(2/3) K^2 $. Therefore we can regard this as a negative cosmological constant. We can even naturally deal with the situation where the the cosmological constant is time dependent since we are free to allow $K$ to change from slice to slice.

Thus  we can consider this matching as gluing a sphere in a maximally sliced anti De Sitter spacetime to a maximal  Schwarzschild exterior. 
This would involve holding $K$ fixed in time.  Alternatively, we could have the interior 
maximal and the exterior be CMC. Obviously, we are only considering the Schwarzschild spacetime in a spherically symmetric slicing. This has no gravitational radiation. Nevertheless, the community of people who numerically analyse binary black hole collisions might find such slicings appealing. 
The black holes could be put in the maximal zone while the radiation could be analysed in 
the asymptotically null CMC zone.

\section{Foliations versus slicings}

If $dK/dt$ is positive and if $dC/dt - dK/dt(R_t^3/3) > 0$, then the lapse function is 
positive everywhere and we have a foliation.  Why is this? How restrictive a condition is it?

If we hold $K$ fixed and increase $C$ we automatically get a foliation \cite{MOM}. 
The lapse is everywhere positive and the slices move forward in time.
However, if we hold $C$ fixed and increase $K$, the slices move forward near both 
infinities and backwards in the middle. One way of seeing this is to look at the (implicit) 
formula for $R_t$,
\begin{equation} \label{R_t}k^2 = 1 - {2M \over R_t} + \left( {KR_t \over 3} - {C \over R_t^2}\right)^2 = 0.\end{equation}
Now compute $dR_t/dK$.
We get
\begin{eqnarray}
\left({M \over R^2} + {K^2 R\over 9} + {KC \over 3R^2} - {2C^2 \over R^5}\right)dR_t  + \nonumber\\
 \left( {KR_t \over 3} - {C \over R_t^2}\right) {dKR_t \over 3} = 0.
\end{eqnarray}
This gives
\begin{equation}
{dR_t \over dK} = {3\left({M \over R^2} + {K^2 R\over 9} + {KC \over 3R^2} - {2C^2 \over R^5}\right) \over
R_t\left( {C \over R_t^2} - {KR_t \over 3}  \right) }> 0.
\end{equation}
This shows that $R_t$ increases as $K$ increases. However, in the upper quadrant 
of the Schwarzschild solution the Schwarzschild radius decreases as one moves 
forward in time from the bifurcation sphere (at $R = 2M$) to the singularity at $R = 0$.
 
\begin{figure}[h]
\begin{center}
\includegraphics[width=60mm]{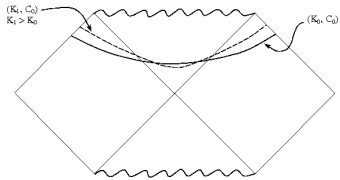}
\end{center}
\caption{  $K$ changes but $C$ is fixed, slices can intersect. }
\end{figure}

Similarly, we have
\begin{equation}
{dR_t \over dC} = -{R_t^2\left({M \over R^2} + {K^2 R\over 9} + {KC \over 3R^2} - {2C^2 \over R^5}\right) \over
R_t\left( {C \over R_t^2} - {KR_t \over 3}  \right) } <  0.
\end{equation}
Therefor $R_t$ decreases with increasing $C$ and so the slice moves forward in time.

Therefore to have the slices moving forward everywhere, when $K$ changes, one needs to 
simultaneously increase $C$ as well as $K$. This behaviour is illustrated in the   two figures.

\begin{figure}[h]
\begin{center}
\includegraphics[width=60mm]{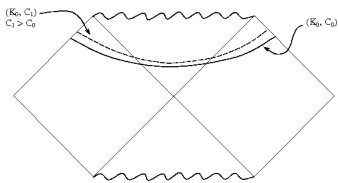}
\end{center} 
\caption{The trace $K$ is constant and $C$ changes; the slices form a foliation.}
\end{figure}
 
One way of finding CMC slices in Schwarzschild is by means of a height function approach, 
see \cite{BOM, MOM}. This involves considering a slice defined as $t = 0$ where $t = T - h(R)$, 
and where $T$ and $R$ are the standard Schwarzschild coordinates. The condition 
that the slice be CMC reduces to a second order differential equation for $h$. 
This can be explicitly integrated once to give \cite{MOM}
\begin{equation}\label{eq:h'}{dh \over dR} = {{KR \over 3} - {C \over R^2} \over \left( 1 - {2M \over R}\right) k}.
\end{equation} 
In retrospect, the fact that the lapse function can be written in such a compact form should
 not have been too surprising. The lapse function can be viewed as the derivative of the height
 function with respect to $K$ and $C$.Therefore one might expect to write the first derivative
 of the lapse as a function, and the lapse itself as a simple integral, just as we do!

We can get an interesting slicing that is `almost' a foliation by setting $C = 8M^3K/3$.
 All these slices have their throats at the bifurcation sphere, $R = 2M$.
 These slices all touch at $R = 2M$, and so the lapse is zero there.  Otherwise, 
as $K$ increases, the lapse is positive everywhere else. This slicing allows $K$ 
to run all the way from $- \infty$ to $+\infty$. If we use $K$ as our label time, 
the lapse function of this slicing is
\begin{eqnarray} \label{eq:Nstar}N =  {1 \over 3}  { 8M^3 - R^3 \over M + {K^2R^3 \over 9} +{8M^3K^2 \over 9}
 - {128M^6 K^2 \over 9R^3}} + \nonumber \\  {k \over 3}\int_{2M}^R {dr \over k}{d \over dr} \left({8M^3 - r^3 \over M + 
{K^2r^3 \over 9} +{8M^3K^2 \over 9} -  {128M^6 K^2 \over 9r^6}}\right)\end{eqnarray}
This slicing covers all of the left and right quadrants, but never enters the upper 
or lower quadrants, the `black' or `white' hole zones.
This shows us that, while we can get a foliation that covers the range 
in $K$ of $[0, \infty)$, we cannot find a foliation that allows $K$ to 
run the whole range $(-\infty, +\infty)$ because the Schwarzschild 
solution is time symmetric and  a foliation that gets to $K = +\infty$ 
must break this. To some degree, this is a word game. If you prespecify 
the range to be covered, i.e., one seeks a foliation that goes from 
$K = -D/M$ to $K = +D/M$, where $D$ is a large number, one can do this. 
Start off with a moment of time symmetry slice , i.e., $(K = 0, C = 0)$ 
and choose a curve in $(K, C)$ space so that $dC/dK$ is only 
infinitesmally bigger than $R_t^3/3$. This will reach any desired value 
of $K$, in particular $K = D/M$. Now add the time reversal of this, and 
we have the desired object. Of course, all such foliations eventually 
run into the critical curve at some finite value of $K$. This, however, 
would be bigger than the specified $D/M$.

\section{Critical foliations}

Henceforth we assume   the subcritical regime --- a minimal surface at $R_t$ and the Neumann
boundary condition $dN/dL = 0$ where $L$  is the proper distance along the slice.  
 This guarantees that the slices are
symmetric about the throat. The line element, expressed in terms of comoving time $t$ and 
the areal radius $R$, takes the
following form
\begin{eqnarray}
ds^2&= & -dt^2\left(  1-{2m\over R} \right) \lambda^2 -\nonumber \\
&&2\lambda {{C\over R^2}-{KR\over 3}\over k }dtdR
+  {dR^2\over k^2} +
R^2d\Omega^2.
\label{explicit}
\end{eqnarray}
Here $\lambda = 1+ \int_{R}^{\infty }dr{\dot C-{r^3\over 3} 
\dot K\over  r^2k^3}$. Let us define following integrals
\begin{eqnarray}
X&=&-\int_{R}^\infty      {    {2\over 3}K^2r^2 +{12C^2\over r^4} 
  \over k\left( 2M + {2Kr^3 \over 9} +{2KC
\over 3} -  {4 C^2 \over r^3}\right)^2} dr  \nonumber\\
Y&=&\int_{R}^\infty {1 \over k}    {r^2\left( 6m +2KC -{24C^2\over r^3}\right)
  \over \left( 2M + {2Kr^3 \over 9} +{2KC
\over 3} -  {4 C^2 \over r^3}\right)^2} dr. 
\label{XY}
\end{eqnarray}
The    formula for the lapse function, $N=k\lambda $, can be rewritten as  
\begin{eqnarray}  
N & = & k+2 { -{R^3 \over 3}  {dK \over dt} + {dC \over dt} 
\over M + {K^2R^3 \over 9} +{KC \over 3} - {2 C^2 \over R^3}}\nonumber \\
 && -{2\over 3} {dK \over dt}k Y
 +2 {dC \over dt} k X .
\label{lapse11}
\end{eqnarray}
This representation of the lapse $N$ is convenient when there exist minimal surfaces.  
 From (\ref{lapse11}) one clearly sees that if the  minimal surface exists at $R_t$,
$k(R_t)=0$, then
\begin{equation}
N=2 { -{R^3_t \over 3}  {\dot K} + {\dot C} 
\over M + {K^2R^3_t \over 9} +{KC \over 3} - {2 C^2 \over R^3_t}}.
\label{lapse22}
\end{equation}
One finds from (\ref{lapse11}) that at the throat
$dN/dL ={1\over 2}{d\over dR}k^2\left( 1+2\dot C X- {2\over 3}\dot K Y\right) $. Therefore the condition $dN/dL = 0$
yields the differential equation
\begin{equation}
\dot C = {-1\over 2X} +{\dot K Y\over 3X};
\label{throat}
\end{equation}
this is a highly nonlinear relation, since both integrals $X$ and $Y$ depend in a convoluted way on $C$ 
and $K$. Nevertheless, it is possible to give a compact analytic description of this foliation
near the critical point $C_*,K_*,R_*$ of the foliation. At this point vanish both $k^2$ and
the first derivative ${dk^2\over dR}$.  The integrals $X$ and $Y$ become divergent at the critical point, 
but the right hand side of (\ref{throat}) is finite everywhere and vanishes at the critical point. Define
\begin{eqnarray}
\Delta &=&{1\over 3}  K^2R^3_t-{3C^2\over R_t^3}+R_t\nonumber\\
\kappa &=&{2\over 3}K^2R^2_t+12{C^2\over R_t^4}.
\label{delta}
\end{eqnarray}
Now assume that $K,C, R_t$ are close to critical values $K_*,C_*, R_*$.
One can show after a lengthy  analysis and a number of careful  estimates, 
that
\begin{eqnarray}
X &\approx &-\sqrt{2\kappa}{R_t\over \Delta^2}   \nonumber\\
Y &\approx &-\sqrt{2\kappa} {R^4_t\over \Delta^2}+ 3\sqrt{2\over \kappa}{R_t^3\over \Delta} 
+\left( {\pi \over 2}-1\right) {72C^2\over R_t\Delta \kappa^{3\over 2}}.
\label{XYX}
\end{eqnarray}

The calculation is  completely analogous to that of Section  VII in \cite{MOM} and it is 
sketched in the Appendix.  
Define $\delta \equiv  R_t-R_*$ and $\epsilon \equiv  C_*-C$. Moreover,
let $\delta K\equiv K_t-K_*$ and $ {\delta K\over \delta }\rightarrow 0$ 
as $R_t$ tends to $R_*$. Then  (up to terms of lower order)
 $\Delta \approx \delta  \kappa $ and $ \epsilon  
 \approx -{B\over A} \delta^2 $. Here 
 $A={R_*^2\over 2}\kappa_*$, $B=-2C_*+{2\over 3} K_*R^3_*$ and $\kappa_*$ is the value of $\kappa $
 at the  critical point.
The insertion of  the above information  into equation (\ref{throat}) and (\ref{XYX}) yields 
the differential equation 
\begin{equation}
{d\over dt} \left( \epsilon -{1\over 3}\delta K R^3_\ast \right) \approx - 
{|B |\kappa^{1/2}  \over \sqrt{2}   R_*^3}  \left( \epsilon -{1\over 3}\delta K R^3_\ast \right) .
\label{20}
\end{equation}
Define $\Gamma ={|B |\kappa^{1/2}  \over \sqrt{2}   R_*^3}$. 
The asymptotic behaviour of the lapse function near the critical point follows from (\ref{lapse22}),
the analysis of the decaying of $k^2$ near $R_*$ 
and (\ref{20}). One obtains
\begin{equation}
N = N_0 e^{-t\Gamma \over 2 }.
\label{24}
\end{equation}
In the particular case of the trace $K$ being independent of time, the estimate (\ref{24}) coincides 
with the result  derived earlier in \cite{MOM}.

\section{Conclusions.}

We consider a family of spherical three dimensional spacelike
 slices embedded in the Schwarzschild solution. The mean curvature is constant
 on each slice but can change from slice to slice. One  describes how the slices
 are stacked by defining the lapse function, that quantifies distance  along the normal as one goes from 
slice to slice.  We write down a simple expression
 for the lapse of any such slicing. This allows us to glue a patch of a Schwarzschild 
spacetime with a CMC slicing to a patch that is maximally sliced. It is easy to identify
 those slicings where the lapse is everywhere positive. The slices do not cross so one has 
a foliation.  There is a barrier that  prevents the mean curvature
 from becoming large, and we show how to avoid this so as to construct a foliation
 where the mean curvature runs all the way from zero to infinity. No foliation exists 
where the mean curvature goes from minus to plus infinity. However, 
if we consider the slicing where each slice passes through the bifurcation sphere, 
the point where $R = 2M$, we almost get a foliation because the lapse only vanishes
 at this one point, and is positive everywhere else, while the mean curvature does 
run from minus to plus infinity. There exist  symmetric foliations of the extended Schwarzschild
 spacetime. They  degenerate at a critical point. We show that the 
 lapse function exponentially approaches  zero at this critical point.

\acknowledgments
This work was supported in part by SFI grant 07/RFP/PHYF148 to N \'O Murchadha.
EM acknowledges support  by the Polish Government  MNII grant 1PO3B 01229 and N \'O Murchadha thanks
Andrzej Sitarz for the financial support within the Transfer of Knowledge programme,  MTKD-CT-2006-042360.
We are grateful to Mark Hannam and Zdobys\l aw \'Swierczy\'nski for their help in preparing figures.

\section{Appendix.}

Below we show how one calculates the integral $X$ defined in the main text.
Define following functions
\begin{eqnarray}
\tilde F_1&\equiv & 6{C^2\over r^4} +{K^2r^2\over 3},
\nonumber\\
{\tilde F_2\over R_t}&\equiv & {K^2R_t^2\over 9} \left( {r^2\over R_t^2} +{r\over R_t} -2\right)
\nonumber\\
&&-{C^2\over R^4_t} \left( {R_t\over r} +{R_t^2\over r^2} +{R^3_t\over r^3}-3 \right) ,
\nonumber\\
\tilde F_3&\equiv &{2\over 9}K^2R^3_t\left( {r^3\over R_t^3}-1 \right) +{4C^2\over R^3_t} \left( 1-{ R_t^3\over r^3}\right) .
 \label{A1}
\end{eqnarray}
The function $k^2$ defined in the main text can be written as follows
\begin{equation}
k^2=\left( 1-{R_t\over r} \right) \left( {\Delta \over R_t} +{\tilde F_2\over R_t}\right)
\label{A2}
\end{equation}
and the integral $X$ takes the form
\begin{eqnarray}
X&=&-2\sqrt{R_t} \int_{R_t}^\infty {dr\over \sqrt{1-{R_t\over r}}} {1\over \sqrt{\Delta +\tilde F_2}}
\nonumber\\
&&
{\tilde F_1\over \left( 2m+ {2\over 3}KC+{2\over 9} K^2r^3-{4C^2\over r^3} \right)^2}.
\label{A3}
\end{eqnarray}
The function $k\left( R\right) $ vanishes at the minimal surface, at $R_t$. Thus 
$2m+{2\over 3}KC =R_t+{1\over 9}K^2R^3_t+{C^2\over R^3_t}$ and the denominator of the third
 factor of (\ref{A3}) can be represented as
\begin{equation}
\left( \Delta +\tilde F_3\right)^2.
\label{A5}
\end{equation}
Here appears the function  $\Delta $, already defined in the main text.
It is now convenient to replace $r$  in the integral $X$ by $y\equiv \sqrt{1-{R_t\over r}}$.
Then $X$ reads
\begin{equation}
X=-4\sqrt{R_t} \int_0^1 dy {F_1\over \sqrt{\Delta +y^2 F_2}\left( \Delta +y^2 F_3\right)^2}.  
\label{A6}
\end{equation}
 Here $F_i \equiv \tilde F_i/y^2$ for $i=2,3$, and $F_1\equiv \tilde F_1/\left( 1-y^2\right)^2$.
Now define a new variable $z\equiv y/\sqrt{\Delta }$; the integral (\ref{A6}) becomes
\begin{equation}
X=-4{\sqrt{R_t}\over \Delta^2} \int_0^{1\over \sqrt{\Delta }}  {dz F_1\left( z\sqrt{\Delta }\right)
 \over \sqrt{ 1+z^2 F_2\left( z\sqrt{\Delta }\right) }\left( 1 +z^2 F_3\left( z\sqrt{\Delta } \right) \right)^2}.  
\label{A7}
\end{equation}
One can split the integral $\int_0^{1/\sqrt{\kappa \delta }}$ into
two parts:  $\int_0^{1/\sqrt{\kappa \delta }}=
\int_0^{1/\sqrt{10^4\kappa \delta }}+
\int_{1/\sqrt{10^4\kappa \delta }}^{1/\sqrt{\kappa \delta }}$.
It is easy to check that the contribution coming from the second
integral of the integrand of $X$ goes to zero as $\delta $ approaches zero. The first
integral of the integrand of $X$ in turn is well approximated by
  the integral
\begin{equation}
X=-4{\sqrt{R_t}\over \Delta^2} \kappa R_t\int_0^{\infty }  {dz \over \sqrt{ 1+{z^2\over 2} R_t\kappa  }
\left( 1 +z^2R_t\kappa  \right)^2}.  
\label{A8}
\end{equation}
This can be explicitly calculated, with the result displayed in the main text.
Let us remark, that   the function $1/\Delta $ explodes to infinity
at the critical point of the foliation. Thus in this limit the quantity $X\Delta^2$
becomes   equal to $-\sqrt{2\kappa_*}$ and the first of equations (\ref{XYX})
appears exact. 
 
The calculation of the other ($Y$) integral defined in Eq. (\ref{XY}) proceeds in a similar way.

\end{document}